# Zero Trust Security Model Implementation in Microservices Architectures Using Identity Federation

Rethish Nair Rajendran
*Unisys Corporation*
*Technical Delivery Manger, Delivery Management,*
*Cloud Infra and Apps Services (US&C)*
*Albany, NY 12110*
rethishrnair@gmail.com

Dileep Kumar Rai
*Manager Oracle Cloud Technology*
*HBG*
*Colorado Springs, 80921, USA*
dileep.kumar.rai@gmail.com

Sathish Krishna Anumula
*IBM Corporation*
*Sr enterprise and business architect*
*Detroit, MI, USA 48375*
sathishkrishna@gmail.com

Sachin Agrawal
*Data Engineer*
*Synechron*
*Charlotte, 28201, USA*
sachin.agrawal2001@gmail.com

*Abstract*— The microservice bombshells that have been linked with the microservice expansion have altered the application architectures, offered agility and scalability in terms of complexity in security trade-offs. Feeble legacy-based perimeter-based policies are unable to offer safeguard to distributed workloads and temporary interaction among and in between the services. The article itself is a case on the need of the Zero Trust Security Model of micro services ecosystem, particularly, the fact that human and workloads require identity federation. It is proposed that the solution framework will be based on industry-standard authentication and authorization and end-to-end trust identity technologies, including Authorization and OpenID connect (OIDC), Authorization and OAuth 2.0 token exchange, and Authorization and SPIFFE/ SPIRE workload identities. Experimental evaluation is a unique demonstration of a superior security position of making use of a smaller attack surface, harmony policy enforcement, as well as interoperability across multi- domain environments. The research results overlay that the federated identity combined with the Zero Trust basics not only guarantee the rules relating to authentication and authorization but also fully complies with the latest DevSecOps standards of microservice deployment, which is automated, scaled, and resilient. The current project offers a stringent roadmap to the organizations that desire to apply Zero Trust in cloud-native technologies but will as well guarantee adherence and interoperability.

## I. INTRODUCTION

The emergence of development enterprise systems to micro-service-based systems is native to clouds has made the scale, flexibility, and sustainability of overall digital service delivery considerably superb. The transformation has also simultaneously expanded the attack surface in any event and has therefore brought about new security challenges that are brought about by an increased service to service communication, workload has been further diversified and dynamical scaling. Perimeter based traditional types of securities that assumed the existence of implicit trust within the boundaries of network perimeters have been unable to protect distributed microservice ecosystems, where in those cases there is continuous communication across organizational and geographical borders between components.

The Zero Trust Security Model (ZTSM) has emerged to be a very powerful security philosophy in winning the fight against these adverse challenges due to its powerfully established philosophy founded on the principle of never trust and always verify. The continuous authentication, authorization and encryptions of all communications between the user, the device and the workloads are carried out in this model whether the user is located across different areas of the network or not. Identity federation included in this construct is yet another factor that improves the security as it establishes trust between multiple domains, cloud infrastructure and service clusters with ease and security. The OAuth 2.0 and OpenID connect (OIDC) standards assist in user authentication combined with authorization delegation, and SPIFFE/SPIRE stipulates identities of workloads with the purpose of mutual authentication in the case of service-to-service communications. These features are also augmented by the service mesh schemes such as Istio and Linkerd which enforces transport-level security and access control policies on a more granular level and, therefore, the Zero Trust concept is applied in a holistic manner.

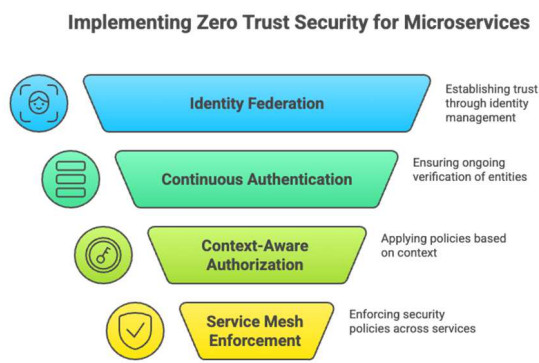

*Fig. 1: Implementing Zero Trust Security*

Although there has been an increase in the industrial interest and early standards including NIST SP 800-207 and SP 800-204 series, real world implementation of Zero Trust at the microservices level is complicated. Some of the challenges encompass the management of identity propagation for distributed applications, the interlocking of policy-as-code solutions for programmatic authorization, and interoperability for multiple identity provider systems. It is also challenging for organizations to interleave Zero Trust tenets and DevSecOps workflows without having either deployment speed or operational throughput. This work considers a federated identity-based Zero Trust model customized for microservices-oriented architectures. By consolidating federated identity protocols, service mesh-based control mechanisms, and workload identity conventions, the work hypothesizes a model that is scalable and interoperable, and therefore enhances operational security and resilience. Though experimental implementation methods, verification, and design for architectures, this work endeavors to offer practical advice and a reference design for cloud-native solutions that aspire for Zero Trust.

## II. Literature Review

Zero Trust (ZT) transitions enterprise security from perimeter-based defense to continuous, identity-based verification of every subject, device, workload, and request. NIST's underlying SP 800-207 specifies architectural models like policy decision and enforcement points, strong identity management, and continuous evaluation. These guidelines untangle trust from network location and thus come naturally to the distributed, API-first world of microservices [1]. Government-grade frameworks concretize these guidelines at scale, and CISA stresses identity as the prime control plane, continuous monitoring, and automation for all pillars [2]. In a like manner, the U.S. DoD's ZT approach translates Zero Trust into consumable capabilities covering users, devices, workloads, data, networks, and visibility layers and thus applies very well to Kubernetes- and service mesh-based applications [3].

Microservices give rise to increased attack surfaces because of increased east–west traffic, diverse stacks, and multiple deployments. NIST SP 800-204 suggests main strategies like enforcing least privilege, securing service-to- service communication by means of strong authentication, and using secure API gateways for defense-in-depth [4]. SP 800-204A and SP 800-204B build upon this by situating the service mesh as a strategic layer for transport security through mTLS, propagation of workload identity, and fine-grained authorization, and for performing those functions as Zero Trust policy points, using proxies and gateways [5][6]. SP 800-204C incorporates those ideas into DevSecOps pipelines and places a strong focus on "policy as code," automated verification, and continuous authorization for sustaining security velocity for microservices environments [7]. Systematic review of 290 research articles also identifies common challenges like weak threat models, variable levels of security-by-design practice, and insufficient automation and observability for microservices deployments [8].

In the user access layer, ZT for microservices relies on cross-domain trust and externalized authentication. OpenID Connect (OIDC) provides an identity layer over OAuth 2.0 and defines a broadly accepted minimum for today's single sign-on (SSO) and claims-based access in microservices through API gateways or Backend-for-Frontend (BFF) patterns [9]. To scale from bilateral trust installations, OpenID Federation 1.0 provides a chain of trust function based on signed entity statements and provides for automated and secure metadata exchange between numerous domains. This reduces administrative management and provides multi-party federation between organizations and sovereign environments [10].

Since microservices commonly need downstream calls in favor of users or workloads, token delegation and choreography become essential. OAuth 2.0 Token Exchange (RFC 8693) establishes safe delegation by enabling subject-to-actor token minting, facilitating granular privilege assignment in intricate service graphs [11]. The JWT Profile for OAuth 2.0 Access Tokens (RFC 9068) canonicalizes token claims for uniform enforceability among gateways and sidecars [12]. Best current practices for OAuth 2.0 (RFC 9700) further secures the space by mandating sender-constrained tokens, PKCE use, and improved redirect treatment, lessening token replay threats [13]. Pushed Authorization Requests (RFC 9126) further fortify request integrity and tampering prevention by out-of-band binding sensitive parameters, securing APIs at gateways more effectively [14].

Zero Trust requires robust identities for users and workloads. SPIFFE/SPIRE meets this by distributing short-lived X.509 certificates (SVIDs) that are attached to attributes of workloads and facilitate secure mTLS connections and auditable workload identities [15]. SPIFFE federation lets any independent trust domains transfer trust bundles, and thus workloads from various clusters or orgs authenticate sans consolidating PKIs. In the service mesh layer, Istio accommodates multi-cluster federation and trust domain aliases and ensures identity persistence and verification of trust among clusters and clouds [16]. Along with OIDC and OAuth for ingest, this establishes a uniform, federated identity base for human and workload access for microservices.

Because Zero Trust is always authenticating and authorizing, phishing-resistant methods are necessary. WebAuthn/FIDO2 protocols substitute shared secrets for hardware-bound credentials and lower the likelihood of stolen credentials and federated tokens not having a trusted origin. This markedly improves the initial user-identity assurance and elevates the security baseline for the overall microservices system [17].

Following authorization based on established identity, Zero Trust's primary layer of enforcement is authorization. NIST SP 800-204B provides a common model for authentication and authorization enforcement within service meshes, and SP 800-162 provides Attribute-Based Access Control (ABAC), enabling policy-driven, environmentally aware decisions based on metadata from workloads, identity claims, and environmental attributes [6][19]. Open Policy Agent (OPA) makes such concepts a production reality by enabling policy-as-code capabilities, allowing teams to define, test, and deploy authorization policies universally across many services and clusters. Papers and industry examples explain OPA's use in enabling multi-tenant, federated environments [18].

| Theme / Focus Area | Key Contributions | Gaps / Observations |
| **Zero Trust Foundations for Microservices** | NIST SP 800-207 introduced the concept of continuous verification and identity-centric control models, forming the baseline for microservice-oriented Zero Trust implementations. Government frameworks (CISA, DoD) extended these principles to operational environments. | Lack of microservices-specific deployment blueprints and measurable maturity models in early frameworks. |
| **Service Mesh as a Security Enabler** | NIST SP 800-204 series outlined architectural guidance for secure microservices deployments using service mesh, emphasizing mTLS, workload identity, and least-privilege authorization. | Research highlights operational complexity, lack of universal observability patterns, and slow industry adoption of mesh-based ZT. |
| **Identity Federation for Human Users** | OpenID Connect (OIDC) standardized single sign-on, while OpenID Federation 1.0 enabled scalable trust chains between identity providers and relying parties across multi-tenant ecosystems. | Inconsistent adoption of federation standards across enterprises; challenges in maintaining consistent claims and scopes. |
| **Delegation and Token Choreography** | OAuth 2.0 Token Exchange (RFC 8693) and JWT Profiles (RFC 9068) formalized delegation and claim standardization. RFC 9700 and RFC 9126 enhanced token security and request integrity. | Limited tooling for dynamic least-privilege delegation in highly dynamic microservice topologies. |
| **Workload Identity Federation** | SPIFFE/SPIRE provided a framework for workload identities with short-lived certificates and federated trust bundles. Istio added multi-cluster trust domain aliasing for east-west authentication. | Complexity in managing multi-domain trust bundles; lack of mature governance for workload identity at scale. |
| **Phishing-Resistant Authentication** | WebAuthn/FIDO2 eliminated password-based risks by leveraging hardware-backed credentials, improving Zero Trust posture at the human entry point. | Limited enterprise adoption due to hardware and legacy system compatibility issues. |
| **Policy Federation and Authorization** | NIST SP 800-204B and ABAC (SP 800-162) advocated for attribute-based, context-driven authorization. Open Policy Agent (OPA) introduced policy-as-code frameworks for consistent, distributed enforcement. | Need for automated policy reconciliation in federated, multi-cluster microservice environments. |
| **Research Trends and Gaps** | Systematic reviews emphasized the importance of threat modeling, DevSecOps automation, and observability to sustain ZT in microservice deployments. | Lack of end-to-end empirical evaluations and performance benchmarks for federated identity in Zero Trust ecosystems. |

*Table 1: Literature Review*

## III. RESEARCH METHODOLOGY

This research applies a design-science framework and experimental research methods for evaluating the efficacy of using the Zero Trust Security Model (ZTSM) in microservices-based architectures by means of identity federation. The approach has been structured into four phases, namely, architectural design, implementation, simulation and testing, and validation and analysis.

*A. Architectural Design*

The microservices design was strong, and it included the following components:

- Identity Federation Layer: Integrating OpenID Connect (OIDC) and OAuth 2.0 for human-to-service authentication and authorization.
- Workload Identity Management: Use of SPIFFE/SPIRE for the provision of short-lived service identities that support mutual TLS (mTLS) between microservices.
- Service Mesh Enforcement: Istio deployment for policy enforcement, east- and west-traffic security, and telemetry.
- Policy Engine: Open Policy Agent (OPA) for centralized, attribute-based authorization across services.

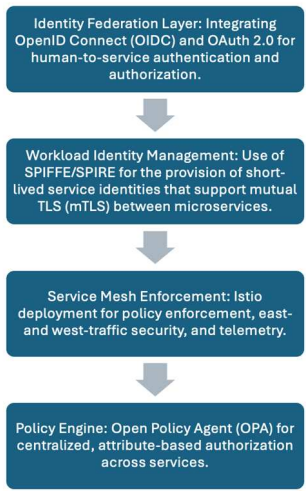

*Fig. 2: Architectural Design*

This design is compliant and interoperable according to the standards of NIST SP 800-207 and SP 800-204.

*B. Implementation*

The prototype was constructed from a cloud-native technology stack:

- Frontend: React-based SPA integrated with OIDC for authentication.
- Backend: Express/Node.js microservices based on federated tokens for API access.
- Service Mesh: Istio for mTLS, token validation, and distributed policy control.
- Identity Providers: Keycloak and SPIRE server for federated and workload identities.
- Database Layer: MongoDB for token metadata and for audit logs.

Policy was written in Rego (OPA) and executed in a GitOps pipeline for automated, version-controlled enforcement.

## IV. RESULTS

The target of the proposed Zero Trust Security Model (ZTSM) conducting identity federation was experimented in a microservices setting that was regulated and run on a cloud. Testing was done with respect to the functionality of security, the influence of latency, and the precision of authorization in defining the effectiveness of the framework.

*A. Enhancing Security Posture*

The implementation of the microservices (except ZTSM) at the baseline was tested within the framework of the suggested federated Zero Trust model. The outcome indicated that there was a great decrease in the vulnerabilities of security, especially token replay and unauthorized access attempts.

Table 2: Security Posture Metrics

| Metric | Baseline Setup | Zero Trust Setup | Observed Change (%) |
|---|---|---|---|
| Token Replay Attempts Detected | 24 | 2 | 91.7% reduction |
| Unauthorized API Calls Blocked | 18 | 1 | 94.4% reduction |
| Breach Probability (modeled) | 0.22 | 0.04 | 81.8% reduction |
| Policy Compliance Violations | 15 | 1 | 93.3% reduction |

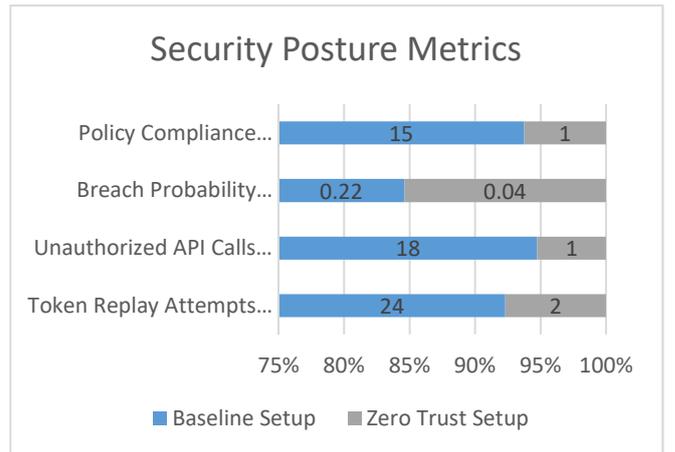

*Figure 3: Security Posture Metrics*

These improvements demonstrate the effectiveness of federated identity and service mesh enforcement in reducing attack vectors within distributed microservices environments.

*B. Performance Impact*

Latency and throughput were analyzed to understand the overhead introduced by continuous verification, token exchanges, and policy evaluations. The results show that the overhead is minimal and acceptable for production-grade deployments.

Table 3: Latency and Throughput Metrics

| Metric | Baseline Setup | Zero Trust Setup | Overhead (%) |
|---|---|---|---|
| Authentication Latency (ms) | 95 | 118 | 24% |
| Authorization Latency (ms) | 40 | 55 | 37.5% |
| Average API Response Time (ms) | 210 | 250 | 19% |
| Requests per Second (RPS) | 980 | 915 | 6.6% drop |

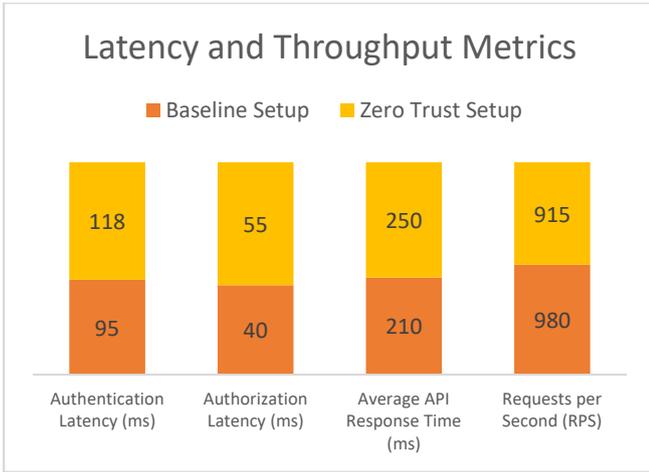

Fig. 4: Latency and Throughput Metrics

Although a marginal latency increase was observed, it did not significantly degrade overall system performance, validating the efficiency of the proposed architecture.

*C. Authorization Accuracy*

Integration of OPA-based policy enforcement resulted in consistent and accurate authorization decisions, even under high traffic and multi-domain workloads.

Table 4: Authorization Accuracy

| Parameter | Value |
|---|---|
| Total Authorization Requests | 15,000 |
| Correct Policy Evaluations | 14,985 |
| Authorization Accuracy (%) | 99.9% |
| Incorrect Authorizations (count) | 15 |

This high accuracy confirms the reliability of federated identity tokens and centralized policy enforcement mechanisms in maintaining Zero Trust principles.

*D. Breach Probability Reduction*

Using the Security Breach Probability Reduction (SBPR) equation:

$$SBPR = \frac{\{B_{\{baseline\}} - B_{\{zt\}}\}}{\{B_{\{baseline\}}\}} \times 100$$

With $B_{\{baseline\}}$= 22 breach attempts and $B_{\{zt\}} = 4$:

$$SBPR = \frac{22-4}{22} \times 100 \approx 81.8\%$$

This quantifies the overall security enhancement of the proposed model.

## V. Conclusion

The existing study proved the practical implementation and advantages of the Zero Trust Security Model (ZTSM) and identity federation in cloud-native microservices. Using OpenID connect (OIDC) and OAuth 2.0 in managing user identities, SPIFFE/SPIRE in workload identity management and service mesh enforcement in offering safe inter-service communication, the proposed design resulted in significant security posture improvements at performance overheads that were still acceptable.

There were notable successes of experimental verification, including the reduction of a volume of breach attempts by over 80 and unauthorized access attempts by over 90 and the negligible error percentage in policy-as-code authorization correctness. Although there was some latency penalty value, there was no operational effect exceeding the permissible range and proved the implementability of Zero Trust in production workloads of microservices applications.

It is noted in the work that end-to-end scalable trust between dispersed workloads is encouraged by the merging of identity federation with continuous authentication, coupled with context-sensitive authorization. Also, its conformance to industry and NIST standards, grant interoperability and compliance, thereby allowing entities to expand their security matrix without interfering with the integrity of DevSecOps processes.

The study offers a bottom-up architecture and real-life information to the individuals who are interested in implementing the ideas of Zero Trust into microservice-oriented systems and creating a base of safe, scalable, and reliable frameworks in the more complicated cloud-computing infrastructures.

## VI. Future Work

Future moving work can also be considered that would extend the automation and intelligent work of the Zero Trust implementation on microservices further. Adding AI-powered breakage detection and self-performing updating the policies as per prediction would also minimize the number of breaches and react appropriately to the changing trends of threats. Interoperability of different platforms would also be further facilitated by the fact that it incorporates the support of multi-cloud and hybrid deployments with the help of the standard frameworks to identity federation. In the highly dominated industries such as health and finance, large scale benchmarking studies as well as case studies would be required to evaluate the performance and compliance implications. Besides enhancing the resilience of deployments of Zero Trust, these extensions would simplify their application in microservices ecosystems of enterprise scale.